\documentclass{elsarticle}

\usepackage[utf8]{inputenc}
\usepackage{lineno,hyperref}
\usepackage[printonlyused,nolist]{acronym}
\usepackage{blindtext}
\usepackage{amsmath}
\usepackage{amssymb}
\usepackage{subfig}
\usepackage{hhline}
\modulolinenumbers[5]

\journal{Future Generation Computer Systems}

\begin{acronym}
	\acro{IDS}{Intrusion Detection System}
	\acro{CIDS}{Collaborative Intrusion Detection System}
	\acro{NIDS}{Network Intrusion Detection System}
	\acro{ML}{Machine Learning}
	\acro{ID2T}{Intrusion Detection Dataset Toolkit}
	\acro{PCAP}{Packet Capture}
	\acro{TTL}{Time To Live}
	\acro{API}{Application Programming Interface}
	\acro{DDoS}{Distributed Denial of Service}
	\acro{DoS}{Denial of Service}
	\acro{TIDED}{Testing Intrusion Detection Datasets}
	\acro{CAIDA}{Cooperative Association for Internet Data Analysis}
	\acro{R2L}{Remote to Local}
	\acro{DARPA}{Defense Advanced Research Projects Agency}
	\acro{MIT}{Massachusetts Institute of Technology}
	\acro{CDX}{Cyber Defense Exercise}
	\acro{NSA}{National Security Agency}
	\acro{USMA}{U.S Military Academy}
	\acro{DNS}{Domain Name System}
	\acro{LBNL}{Lawrence Berkeley National Laboratory}
	\acro{KDD}{Knowledge Discovery and Data Mining}
	\acro{SIGKDD}{Special Interest Group on Knowledge Discovery and Data Mining}
	\acro{MAWI}{Measurement and Analysis on the WIDE Internet}
	\acro{r2l}{Remote to Local}
	\acro{u2r}{User to Root}
	\acro{WIDE}{Widely Integrated Distributed Environment}
	\acro{USMA}{U.S Military Academy}
	\acro{IMPACT}{Information Marketplace for Policy and Analysis of Cyber-risk \& Trust}
	\acro{UMass}{University of MAssachusetts Amherst}
	\acro{ADFALD}{Australian Defence Force Academy Linux Dataset}
	\acro{IRSC}{Indian River State College}
	\acro{UNSWNB}{University of New South Wale Network Based}
	\acro{FLAME}{Flow-Level Anomaly Modeling Engine}
	\acro{IRC}{Internet Relay Chat}
	\acro{NERDS}{Network Entity Reputation Database System}
	\acro{CTI}{Cyber Threat Intelligence}
	\acro{FMP}{Future Misbehavior Probability}
	\acro{ASN}{Autonomous System Number}
	\acro{ML}{Machine Learning}
	\acro{EWMA}{Exponentially Weighted Moving Average}
	\acro{BS}{Brier Score}
\end{acronym}

\begin{document}


\begin{frontmatter}

\title{Network entity characterization and attack prediction} 

\author[cesnet]{Vaclav Bartos\corref{correspondingauthor}}
\ead{bartos@cesnet.cz}
\author[cesnet]{Martin Zadnik}
\ead{zadnik@cesnet.cz}

\author[tuda]{Sheikh Mahbub Habib}
\ead{sheikh.mahbub.habib@continental.com}
\author[aau]{Emmanouil Vasilomanolakis}
\ead{emv@cmi.aau.dk}

\cortext[correspondingauthor]{Corresponding author}
\address[cesnet]{CESNET, Czech Republic}
\address[tuda]{Continental AG, Germany}
\address[aau]{Center for Communication, Media and Information Technologies, Aalborg University}

\begin{abstract}
The devastating effects of cyber-attacks, highlight the need for novel attack detection and prevention techniques. Over the last years, considerable work has been done in the areas of attack detection as well as in collaborative defense. However, an analysis of the state of the art suggests that many challenges exist in prioritizing alert data and in studying the relation between a recently discovered attack and the probability of it occurring again.  
In this article, we propose a system that is intended for characterizing network entities and the likelihood that they will behave maliciously in the future. Our system, namely \acf{NERDS}, takes into account all the available information regarding a network entity (\emph{e.\,g.} IP address) to calculate the probability that it will act maliciously. The latter part is achieved via the utilization of machine learning.
Our experimental results show that 
it is indeed possible to precisely estimate the probability of future attacks from each entity using information about its previous malicious behavior and other characteristics.
Ranking the entities by this probability has practical applications in alert prioritization, assembly of highly effective blacklists of a limited length and other use cases.
\end{abstract}


\end{frontmatter}



\section{Introduction}
With cyber-attacks increasing both in numbers and sophistication, a lot of research has been done over the last years towards collaborative detection mechanisms. Such research varies from sophisticated alert data correlation and aggregation mechanisms to the construction of complex collaborative architectures \cite{vasilomanolakis2015survey}.
As a result, a plethora of alert sharing platforms and systems have been proposed\footnote{Note that the state of the art utilizes a multitude of different terms to describe semantically similar systems. Such terms include but are not limited to: \acp{CIDS}, collaborative intrusion detection networks, threat intelligence sharing platforms, network telescopes, etc.}.
Nevertheless, a more in-depth analysis of threat sharing platforms and systems shows that a number of challenges need to be addressed before such systems can be considered mature \cite{vasilomanolakis2015survey,meng2015collaborative}.

First, we argue that one of the core issues is the large number of alerts, generated by the cybersecurity tools, an analyst deals with. 
This issue is amplified when considering additional data received via various sharing and collaborative platforms.
In this context, data prioritization and summarization are essential for reducing the overwhelming amount of information presented to the analyst.
Indeed, prioritization was identified as one of the most important parts of cyber incident handling processes in several studies~\cite{ENISA2010, ponemon2018}.
   

Second, the reasoning behind exchanging alert data (as well as blacklisting) usually comes with an implicit assumption that recently discovered attacks are likely to be performed again in a similar manner or by the same attacker. 
However, this holds only for a certain set of attackers and attack types, while others appear to be non-repetitive or even one time only.
The analyst needs to be able to effectively recognize in which set an attacker belongs to, in order to initiate further actions that are relevant to the given set.
For instance, the identification of a persistent attacker leads to its automated blocking, while an one-time-only attack, on a critical asset, leads to further investigation.

Third, there are numerous blacklists and other threat intelligence sources as well as multiple alerts from security monitoring tools. 
Not only the volume of information is high but some data are irrelevant or of low quality.
While collecting all relevant data is important for a detailed attack analysis, in other cases it is important to be able to quickly comprehend the main properties of the attacking source, so it is possible to easily or even automatically assess the source behavior and decide about the appropriate immediate action.

We argue that a well designed method of summarizing all known information about a malicious network entity can help to address the aforementioned issues.
%
To this end, in this article we propose a machine learning based algorithm to estimate the probability that a particular entity (e.g. an IP address)
will repeat an attack in the near future.
%
We call this probability estimation the \textit{\acf{FMP} score}.
The score represents an aggregated knowledge about each entity and it expresses its expected behavior; allowing also for the comparison between entities.
The previous works on scoring IP addresses or networks lack some important properties (such as prediction of the future behavior), only allow the assessment of whole networks, or cannot use some important input features.

The score is a key enabler for several network security applications.
Network administrators can utilize it to prioritize the alert data they receive.
For example, it is common to utilize alert sharing platforms to be alerted on malicious hosts in the network.
In such a use case, the administrator watches the alerts shared by others and if the reported IP address belongs to her constituency,
it indicates there is a misbehaving (e.g. malware infected) host in the managed network.
Hence, the score supports the administrator's decision process (i.e. which IP address to deal with first), especially in large constituencies.

Besides alert prioritization, the score has practical applications in attack mitigation and traffic analysis.
A straightforward usage is to assemble entities with the highest \ac{FMP} score into a blacklist, which is then used to block network traffic from these entities.
Furthermore, the \ac{FMP} score may serve indirectly as one of the decision criteria in spam filters, DDoS mitigation devices or any other algorithms recognizing malicious traffic by multiple criteria.
In addition, the existence of an \ac{FMP} score also offers new possibilities of traffic analysis.
For example an \ac{IDS} may apply more detailed analysis techniques (which would not scale for all the traffic) to the traffic belonging to the highly-ranked entities.



At a glance, the main contributions of this paper are as follows:

\begin{itemize}
  \item We introduce the concept of an advanced reputation database system for network entities to improve alert processing and prioritization. 
  \item We propose a generic method for ranking network entities by a \textit{\acf{FMP} score}, a value that summarizes all known information about an entity to express the level of threat it poses. 
  \item We evaluate and compare different machine learning approaches and sets of input features to identify the most efficient ones, with regard to the specific application scenario of ranking malicious IP addresses. We also evaluate how the FMP score can be utilized for creating predictive blacklists.
\end{itemize}

The evaluation is based on millions of alerts from a real alert-sharing system.
The results show that the proposed method indeed creates a predictor which is able to accurately estimate the probability of future attacks per each evaluated IP address.
Moreover, the predictor assigns scores smoothly over the whole range between 0 and 1, rendering it well usable for ranking addresses 
and prioritization.
Our evaluation also demonstrates the advantage of having the score assigned to each known malicious IP address when a blacklist of a limited size needs to be created.
The FMP score allows to create the most effective blacklist possible for any given size.

The remainder of this article is structured as follows.
Section \ref{sec:related} discusses the state of the art 
in threat intelligence platforms and in methods of evaluating reputation of network entities.
Section \ref{sec:overview} provides context to our scoring method by presenting a general model of an alert processing system and introducing the reputation database system NERDS.
In Section \ref{sec:repscore}, we describe in detail the generic FMP score estimation algorithm.
Subsequently, Section \ref{sec:evaluation} provides the evaluation of the scoring mechanism along with a comparison of different machine learning models and evaluation of one of the possible uses of the score -- generation of predictive blacklists.
Lastly, Section \ref{sec:conclusion} concludes this article and outlines our future work plans.

\section{Related work}
\label{sec:related}

Our work is related to existing threat intelligence platforms.
A brief overview of the platforms is provided in the following subsection.
Subsequently, we compare our work with recent research on  
various characteristics of malicious sources and with the state-of-the-art 
approaches for evaluating reputation in the network domain.


%


\subsection{Threat intelligence platforms}

Many platforms exist for cyber threat intelligence management and sharing -- both open and free as well as commercial ones.
Examples of open platforms for exchanging data about cyber threats and indicators of compromise are MISP~\cite{wagner2016misp}, Warden~\cite{warden, warden2016}, DShield\footnote{\url{https://www.dshield.org/}} and CIF\footnote{\url{https://csirtgadgets.com/collective-intelligence-framework}}.

The Malware Information Sharing Platform (MISP) is an open source solution for collection, storage, distribution and sharing of indicators and alerts regarding cyber security incidents. The main goal of MISP is to share information related to targeted attacks and malware. The features include a centralized searchable data repository, a sharing mechanism based on defined trust groups, and semi-anonymized discussion boards.

The Warden is an open source platform designed for automated sharing of detected security events.
It enables CERTs/CSIRTs (and security teams in principle) to share as well as make use of information on detected attacks and anomalies in networks or services as generated by different detectors -- intrusion detection system (IDS), honeypots, network probes, traffic logs, etc.

DShield is a platform for collection and analysis of incoming malicious activities detected by thousands of contributors.
The contributing network operators send alerts from packet filters like firewalls or IDS systems,
DShield aggregates and analyzes them and provides various statistics as well as blacklists of the most dangerous networks on its website.

The Collective Intelligence Framework (CIF) is a cyber threat intelligence management system, which allows to collect data, mostly indicators like IP addresses, FQDNs and URLs, from multiple sources. It allows to parse, normalize, store, post process and query them and also to share them to others. 

A list of commercial threat intelligence sharing and management platforms includes, for example,
SoltraEdge, IBM X-Force Exchange, Facebook Threat Exchange, Alien Vault Open Threat Exchange
and many others~\cite{sauerwein2017threat}.

The term cyber threat intelligence is usually understood as a way to encompass high-level information, such as description of techniques and procedures used by adversaries, information about malware or phishing campaigns, or global trends in different types of threats. Nevertheless, most of the existing tools focus primarily on sharing of indicators of compromise, such as malicious IP addresses, URLs or file hashes~\cite{sauerwein2017threat}.
A recent user study~\cite{ponemon2018} also agrees that despite the significant growth of threat intelligence platforms in the last few years, the most often used source of threat data are still classic blacklists (mostly lists of malicious IP addresses and URLs).

These blacklists represent a trivial way to express reputation of network entities. Although they are easy to use they do not offer granularity and details.
The information provided by blacklists is only binary -- an entity is either listed or not.
There are usually no details describing the type, intensity or frequency of malicious activities performed by the entity nor any score evaluating the level of threat the entity poses or the confidence of its listing, which would allow to sort or filter them.
Such information would be very useful.
Multiple studies~\cite{ponemon2018,ENISA2010} recognized the ability to prioritize threats and the ability to get a comprehensive picture of the threat among the most important requirements on threat intelligence.
In this context, this article focuses on collecting detailed information about the malicious network entities and on their scoring.

\subsection{Evaluating reputation of network entities}


This work builds on the knowledge of various characteristics of malicious traffic sources.
These characteristics has been studied in several previous works, mostly in the context of IP addresses.
In one such work Collins et al.~\cite{collins2007uncleanliness} show that devices in some networks are more prone to be compromised (e.g. infected with malware) and cleaning up those devices takes longer than in other networks\footnote{Networks in this work are defined simply as IP prefixes of length /24.}.
This network property is called \emph{uncleanliness} and the authors propose a method to quantify it using data about known malicious IP addresses from different sources.
The measures of spatial and temporal uncleanliness of different networks are then used to predict which networks are likely to contain bots or otherwise malicious addresses.

Shue et al.~\cite{shue2012abnormally} presented a similar work which focuses on the analysis of malicious autonomous systems (AS).
By using data from several blacklists as well as their own spam detection tools, authors show that some AS contain much more malicious IP addresses than others.
While in most cases it is probably caused just by poor security policies in those networks, there are also AS with more than 80\,\% of their address space blacklisted. 
Authors argue that these AS are probably run only to host malicious activities. 

The non-uniform distribution of malicious sources in the IP address space was also studied in a series of works on the so called \emph{bad neighborhoods}~\cite{wanrooij2010filtering,moura2011spam,moura2012aggregation,moura2013phd}, which is the term used for networks with high ratio of malicious IP addresses.
The authors propose to aggregate IP addresses listed on various blacklists by their common prefix (usually of length /24)
and create lists of prefixes (networks) with too many blacklisted IP addresses.
Such \emph{bad neighborhood blacklists} can then be used in spam filtering algorithms.
In these works, authors also analyze various characteristics of the bad neighborhoods.
For example, in~\cite{moura2011spam} they show that in case of spam sources, only 10\,\% of the most active neighborhoods (/24 prefixes) are responsible for more than half of all the spam messages.
In other works they show that the existence of bad neighborhoods can also be observed in data about other types of malicious activity, like attacks on SSH, but that the particular lists of malicious networks are different for each type of such activity.

In a later work~\cite{moura2014temporal} Moura et al. undertook research into temporal characteristics of the bad neighborhoods.
Using several datasets about different types of malicious activities, they found out that 40--95\,\% (depending on the dataset) of bad neighborhoods repeat an attack against the same target within a few days.

All the aforementioned works analyze the maliciousness of the whole networks, usually defined as groups of IP addresses with the same /24 prefix.
Some of them propose a score to rate the networks.
Such scoring methods are usually based on the number of blacklisted hosts in a network.
Such an approach is simple as it utilizes the spatial correlations among the malicious IP addresses.
However, the scoring of the whole networks represents certainly a significant issue and limits its applicability. 
For example, in case of blocking the whole network prefix many completely benign addresses are blocked.

There are also several works studying properties of malicious traffic sources at the level of individual IP addresses. Zhang et al. \cite{zhang2013characterization} take 9 public blacklists and analyze both temporal and spatial characteristics of their entries. They show, for example, that the lists are changing quickly and that even the geographic distribution of malicious IP addresses around the world is highly non-uniform.
Another characteristics are shown in \cite{thomas2016abuse}, where authors analyze lists of IP addresses reported as malicious by various Google services. For example, they show that 1\,\% of the most active malign IP addresses are responsible for 48--82\,\% of all attacks (depending on the service attacked). They also found significant correlations between lists of addresses attacking different services, i.e. in some cases a single address is used to attack multiple services.
Similar characteristics of behavior of malicious IP addresses are observed in other works, such as Wahid's work~\cite{wahid2013estimating} or in our own previous work~\cite{warden-analysis,alert-analysis-camad}.

Another set of works that inspired our method are those on the topic of \emph{predictive blacklisting}~\cite{zhang2008hpb, soldo2011blacklisting}.
These works propose methods to explicitly predict which malicious sources are likely to attack in a near future
(in contrast to classic blacklists, which only list those attacking in the past).
The goal of the proposed methods is to prepare a blacklist for each organization which contains those sources that are the most likely to attack the organization network within the next day.
Zhang et al.~\cite{zhang2008hpb} introduced the concept of creating these targeted \textit{highly predictive blacklists} 
and provided a method based on leveraging correlations among sets of attackers targeting individual organizations.
Subsequently, Soldo et al.~\cite{soldo2011blacklisting} presented a method that significantly improved the precision of generated blacklists. It models the problem as a recommendation system which combines several prediction methods.

Although the methods internally work with a ranking of attackers by their probability of attacking the given target network,
the only goal is to build a blacklist of a predefined size (as a top-n list of such ranked attackers).
Therefore, there is no evaluation by a score with a well defined meaning.
Moreover, the attackers in these works are always whole /24 prefixes, not individual IP addresses, so the same disadvantages stemming from low granularity, as those described for network scoring methods above, apply as well.

A recently published paper~\cite{misp-indicator-scoring} describes a work in progress on a method for scoring individual IP addresses and other identifiers (so called indicators of compromise) in the context of the MISP threat sharing platform.
The score is used to estimate whether an indicator is still relevant or not.
It is based on indicator observations, assigned tags and reliability of data sources.
The score of an indicator is reset to its maximum value every time an observation of that indicator in the wild is reported and it decreases in time by a predefined formula. When the score reaches zero, the indicator is marked as expired and can be discarded.
However, the work is still quite incomplete. For example, the meaning of values between the maximum and zero is not defined, and the method for evaluating source reliability has not been designed yet.
Also, spatial correlations among the neighboring IP addresses are not used in any way and there is no attempt of explicit prediction of the future behavior.


\begin{table}[t]
\centering
\caption{Comparison of previous methods for evaluating reputation of network entities.}
\label{tab:related-works-summary}
\begin{tabular}{|l|c|c|c|c|c|} \hline
    & granularity & numerical & predictive & neighbor- & other \\
    &             & score     &            & hoods     & data  \\\hline
Traditional blacklists
    & yes         & no              & no         & no            & no \\\hline
Scoring of networks \cite{collins2007uncleanliness,wanrooij2010filtering,shue2012abnormally}
    & no          & yes             & no         & yes           & no \\\hline
Predictive blacklisting \cite{zhang2008hpb,soldo2011blacklisting}
    & no          & no              & yes        & yes           & no \\\hline
MISP scoring \cite{misp-indicator-scoring}
    & yes         & yes             & no         & no            & no \\\hline
FMP score (this work)
    & yes         & yes             & yes        & yes           & yes \\\hline
\end{tabular}
\end{table}


Table~\ref{tab:related-works-summary} summarizes important aspects of the previous methods for evaluating reputation in contrast to our FMP score.
The \emph{high granularity} column shows if the score is assigned to each individual IP address, not just network prefix or other large group of addresses.
\emph{Numerical score} expresses if there is some numerical value assigned to each entity, or it is just a simple list providing only binary information about the entities (listed or not listed).
The \emph{predictive} column depicts whether the method is based on explicit prediction on future behavior of the evaluated entities or not.
The \emph{neighborhood} column shows whether the method utilizes the correlations among neighboring IP addresses or not.
The last column shows if the method is able to utilize any other data than those regarding previously reported malicious activities.

The table shows that each of the previous works lacks some of these aspects.
Our proposed FMP score is the first non-binary evaluation of reputation of individual network entities which is based on prediction of future attacks.
For the prediction, we consider not only the previous behavior of the evaluated entity and its neighboring entities,
but also other related information not directly derived from observed behavior.

\section{\acf{NERDS}}\label{sec:overview}



\begin{figure}[b!]
	\centering
	\includegraphics[width=1.0\textwidth]{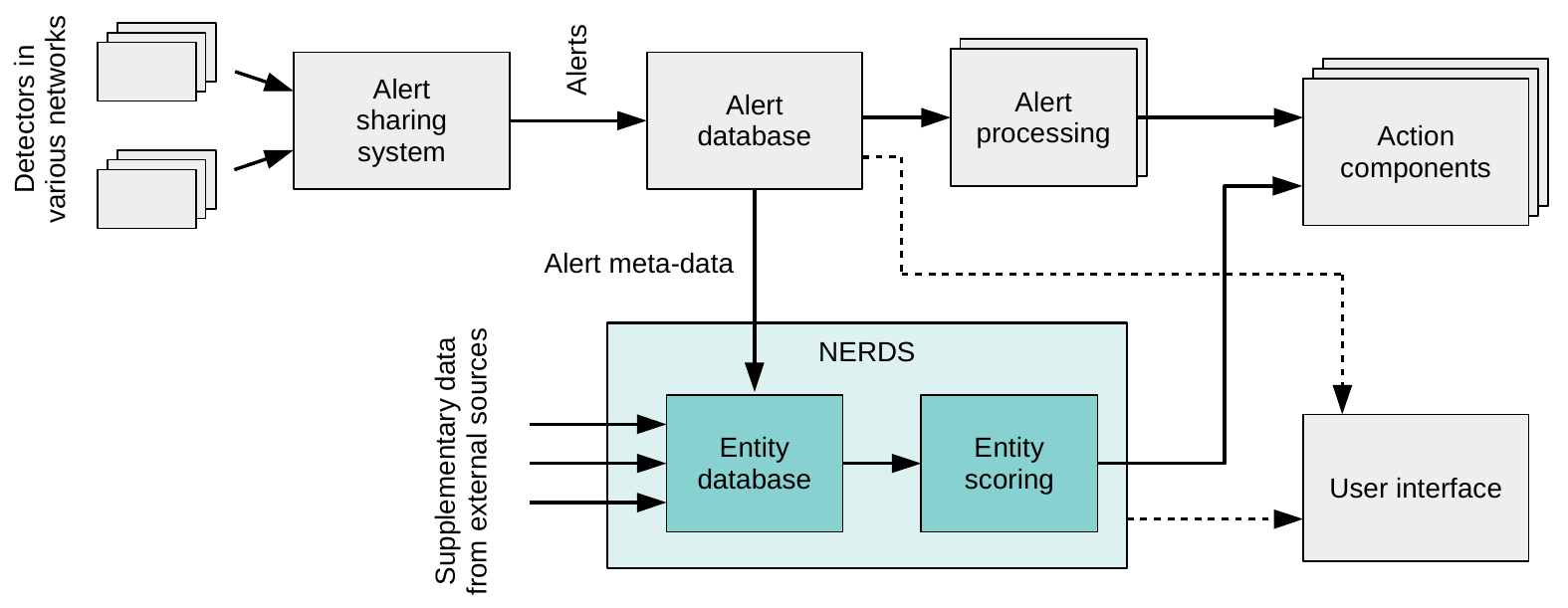}
	\caption{Conceptual framework of an alert processing system with entity database}
	\label{fig:systemarch}
\end{figure}

To provide context to our work, this section briefly introduces \ac{NERDS} (Network Entity Reputation Database System).
\ac{NERDS} stores information about malicious network entities and summarizes the pieces of information into the FMP score. 
\ac{NERDS} is a component of a larger \ac{CTI} infrastructure.
As a particular example of the \ac{CTI} infrastructure, we consider an ecosystem of components which collect, analyze and react upon network alerts depicted on Figure~\ref{fig:systemarch}.

The network alerts are collected by an alert sharing component from a plethora of network monitoring mechanisms, capable of detecting and reporting malicious network activities. For instance, from honeypots, \acp{IDS}, network behavior analyzers, log analyzers, etc.
We utilize the alert sharing unit as an example input component, since alert sharing systems are becoming popular, widely deployed, as well as they usually share a large amount of alerts from diverse sources. At the same time they provide data normalization. 
Nevertheless, the alert sharing can be complemented or replaced by any other collection component without affecting the presented approach (provided sufficient number of alerts is collected to allow for effective application of machine learning techniques).
The collected alerts are stored in an alert database.
The database offers a fast query interface to mine data in the stored alerts either by a user from a user interface or automatically by an analysis component.
The analysis component extracts relevant information out of the alerts.
Analysis results are displayed to an expert (e.g. CSIRT/CERT) via a user interface, moreover, some results are transformed into actions in the Action component such as trigger a \ac{DDoS} protection, issue an incident, notify an operator. 

While some pieces of information are straightforward to obtain from alerts, for example by filtering and aggregation, some remain hidden deeper and a more sophisticated approach must be applied to receive satisfying results.
We consider \ac{NERDS} as such an advanced analysis approach as it gathers available knowledge about the history of observed network entities and predict their future behavior based on this knowledge.

In more details, \ac{NERDS} consists of two parts -- the \textit{entity database} and the \textit{entity scoring}.
The entity database keeps a record for each entity (e.g. IP address) reported as malicious by one or more of the alerts.
The record not only contains meta-data about the observed alerts but also additional relevant information from various external sources to broaden the visibility of the behavior of the entity in a more global scale.
In case of IP addresses, the information kept in records include, for example, resolved hostname, geo-location, autonomous system number or occurrence of the IP address on several public blacklists.

While the entity database is a necessary prerequisite the entity scoring is the core part of the advanced analysis.
The entity scoring mechanism summarizes all available information gathered per  entity into a score and this score is assigned per entity.
The score represents a meaningful and actionable information that is utilized by the action components, for example, to block traffic from most offending IP addresses or domains, or by a user directly, for example, to prioritize investigation of reported incidents or to bring attention to a prevailing issue. A first idea of such a reputation database, including summarizing the data into a single reputation score, has been briefly introduced in our earlier work~\cite{reputation-aims16}.
In this work we propose and evaluate a particular method which can be used in the scoring component.

\section{\acf{FMP} Estimation}\label{sec:repscore}

One important element of NERDS is the scoring component, which estimates the entity score based on all the stored knowledge about an entity.
Since we believe that a well-understood mechanism increases trust and utilization of the score in real-world applications, we devote this section to a thorough description of the estimation mechanism.

While our main motivation is scoring individual IP addresses, the formal description of the method provided below is general to account for 
any kind of network identifier that may be reported as malicious, e.g. an IP address, a domain name, an autonomous system, etc.
We narrow the generic concept into a method for scoring IPv4 addresses in Section~\ref{sec:evaluation}.

\subsection{General Concept}

At a glance, the input of the scoring component consists of two types of data,
(\textit{i}) meta-data about the reported alerts related to the entity (and optionally to its close neighborhood as well),
and (\textit{ii}) complementary, relevant third-party security information related to the entity.
Such complementary information includes, for example,
the hostname, the \ac{ASN} or the geo-location data of IP addresses,
the domain name entropy or the date of registration for domain names, information about the presence of an entity on public blacklists, etc.
Both data types are stored and provided by the Entity database which periodically gathers it from a multitude of sources.

An output of the scoring component is a number, which expresses the level of threat (or maliciousness) the entity poses.
We coin this as {\em \acf{FMP} score} 
%
%
and we define the \ac{FMP} score of an entity as {\em the estimated probability that the entity will behave maliciously in an upcoming time interval (prediction window)}.

The score is, therefore, a result of prediction of future malicious activities 
of the entities based on all the available information about the entity.
We propose to use a \ac{ML} technique to derive a predictor since (\textit{i}) the amount of available information is large, (\textit{ii}) the prediction model is not straightforward to derive analytically and (\textit{iii}) the predictor is expected to be periodically adjusted to the characteristics of the latest data.

An ideal predictor, capable of predicting the future precisely, would only assign \ac{FMP} score of $1.0$ or $0.0$, depending on whether the entity behaves maliciously or not in the prediction window.
However, it is impossible to assemble such a predictor in practice. That is, any real-world predictor is only able to estimate the probability based on information available at the time of prediction.
Therefore, our goal is to design a high precision predictor, by the means of minimizing the error of estimated probability over all entities.

Note, that in practice it is usually impossible to find out all malicious behavior of an entity, since the predictor receives only the detected ones (via alerts received from detection systems). As a result, it is only able to predict future \textit{alerts} related to the entity and not all actual attacks. The quality of the input data in terms of accuracy and coverage has impact on the quality of prediction. The better the input data, the more precise and useful is the \ac{FMP} score. Nevertheless, the principle of the scoring method is robust enough to work with low quality data as well.

The \ac{FMP} score may be general, predicting any kind of malicious network behavior, or specific to a particular type of activity.
For example, there may be an \ac{FMP} score in the context of \ac{DDoS} attacks and a different one in the context of port scans, each estimating probability of different types of behavior.
It is also possible to compute the \ac{FMP} score for specific targets, e.g. for specific networks or types of services.
When it is needed to distinguish multiple such FMP scores, an index is used, e.g. $\mathit{FMP}_\text{scan}$.
In the rest of this section, we will not differentiate between these variants, since the only difference is in what is considered a malicious behavior that should be predicted.


The length of the prediction window should conform to the particular application use case.
The longer time preference the application has the longer the window. There can be applications requiring both short and long-term expectation of entity's behavior separately, thus multiple windows lengths must be predicted in parallel.
In this work, we consider $24$-hour prediction window as a medium length that satisfies majority of applications
(which is also in line with the previous works~\cite{zhang2008hpb,soldo2011blacklisting}).

\subsection{Formal definition}
\label{sec:formal}

The main input of the method are alerts reporting malicious activity of some entities.
Alerts may have different formats and contain various information, but for the  purpose of our work, 
we assume that each alert contains at least: 
(\textit{i}) time of detection, $t$,
and (\textit{ii}) identification of the entity reported as source of the event\footnote{
In case an alert contains multiple sources it can be replaced by multiple alerts with a single source and equally distributed volume.}
(e.g. source IP address), $e$.
Preferably, it should also contain:
(\textit{iii}) type or category of the event, $c$,
(\textit{iv}) event volume, $v$ (its exact meaning depends on the event type, e.g. a number of connection attempts), and
(\textit{v}) identification of the detector, $d$.
In the following text, we assume that alerts contain all these five attributes,
but the method may be applied, with some limitations, utilizing only the first two as well.

An alert can therefore be defined as a tuple $a = (t, e, c, v, d)$.
A set of all alerts available is denoted as $A$.
The time at which the prediction is computed (\textit{current time} or \textit{prediction time}) is marked as $t_0$.
The \textit{prediction window}, $T_p$, is the time window of length $w_p$ immediately following $t_0$, $T_p = (t_0, t_0 + w_p)$.
The predictor uses information about the past alerts from a \textit{history window}, $T_h = (t_0 - w_h, t_0)$, where $w_h$ is the history window length.

For a given prediction time $t_0$, a feature vector $\mathbf{x}_{e,t_0} = (x_1, x_2, ..., x_k)_{e,t_0}$ is computed for each entity $e$.
The feature vector consists of various alert-based features, computed from alerts received within the history window,
and the non-alert features, extracted from other available information related to the entity at $t_0$
(see Section \ref{sec:feature-selection} for further discussion of features).

The output to be predicted (class label), $y_{e,t_0}$, is binary; depicting whether or not there is an alert reporting the entity within the prediction window:

\begin{equation}
y_{e,t_0} = 
\begin{cases}
1 & \text{if} ~ \exists a \in A: a = (t, e, \cdot, \cdot, \cdot), t \in T_p \\
0 & \text{otherwise}
\end{cases}
\end{equation}

If an \ac{FMP} score in some context is to be computed, the condition above becomes more restrictive, e.g. the alert category must match a given value.
Samples with $y_{e,t_0} = 1$ are said to belong to the \textit{positive class},
the others form the \textit{negative class}.

Now, the task is to create an estimator which, for a given feature vector $\mathbf{x}_{e,t_0}$,
is able to accurately estimate the probability that $y_{e,t_0} = 1$,
i.e. that the entity will be reported as malicious in the prediction window.
This task is known as \textit{binary class probability estimation problem} in the machine learning community.
That is basically a binary classification problem where we are not interested in final class assignments but rather in the probability of each class.

Output of the estimator is denoted as $\hat{y}_{e,t_0}$ and represents the estimated probability of the positive class given the feature vector,

\begin{equation}
\hat{y}_{e,t_0} \approx p(y_{e,t_0} = 1 | \mathbf{x}_{e,t_0}).
\end{equation}

To create the estimator we follow the common supervised machine learning process.
First, we create an annotated dataset.
Each \emph{sample} in the dataset describes a particular entity at a particular time.
We select one or more time instances in which the features are computed.
We denote the set of these \textit{sample times} as $T_s = {t_1, ..., t_m}$.
A pair of feature vector and class label, $(\mathbf{x}_{e,t_0}, y_{e,t_0})$,
is then computed for each entity $e \in E$ at each sample time $t_0 \in T_s$,
creating a dataset of $|E \times T_s|$ samples.
From now on, we will index the samples of the dataset by $i$ for more concise notation.

The dataset is then randomly split into a training and a testing set and
the first one is used to train the model\footnote{
We do not recommend any particular model in this generic part. Usually multiple models with various configurations need to be tested and the one performing the best in the particular application is chosen.}.

The metric suitable for training and evaluating the model in this type of problem is the \ac{BS}.
In our binary case, with classes labeled $0$ and $1$,
the \ac{BS} can be described as a mean squared difference of 
predicted probability of the positive class and value of the real class:

\begin{equation}
\mathit{BS} = \frac{1}{N} \sum_{i=1}^N\limits (\hat{y}_i - y_i)^2,
\end{equation}

where $N$ is the number of samples.
The \ac{BS} takes values between 0 and 1. Lower values signify a more accurate prediction.

After the model is trained and its performance is acceptable, it is used to assign \ac{FMP} scores to new samples of network entities.
For each new entity to be evaluated, a feature vector is computed from all related alerts and other available information,
and it is passed to the trained model. Its output, $\hat{y}$, is then directly used as the \ac{FMP} score,

\begin{equation}
\mathit{FMP}(e,t_0) = \hat{y}_{e,t_0}
\end{equation}

A change in behavior of malicious entities as well as in the configuration of detectors influence the characteristics of alerts.
Therefore, the model should be re-trained on new data whenever the performance of the predictor decreases below a defined threshold.

If multiple \ac{FMP} scores for different contexts are required (e.g. for different types of network attacks), a separate model must be trained for each such context using different training data.
Samples in the training data are labeled as positive class ($y_i = 1$) only when there is an alert of given type in the prediction window, alerts of other types are ignored (samples have $y_i = 0$). Nevertheless, the input features still contain information about all types, since there may be correlations between different attack types that can be exploited by the predictor (e.g. login attempts are often preceded by port scans~\cite{ssh-compromise,alert-analysis-camad}, so information about scans can be used to improve prediction accuracy of login attempts). Of course, information about different attack types should be treated as different features.

\subsection{Feature selection guidelines}
\label{sec:feature-selection}

As already noted, a feature vector for our scoring method generally consists of two types of features:
(\textit{i}) the features based on previous alerts related to the entity or similar entities (e.g. nearby IP addresses), and
(\textit{ii}) the features based on other data sources than alerts (e.g. presence of the entity on a public blacklist).

The particular set of features needs to be designed specifically for each class of entities and according to the input data available.
However, at least for the alert-related features, we provide basic guidelines and examples that we expect to work well in most cases.
We propose the utilization of the following characteristics as the basis for alert related features:

\begin{itemize}
\item Number of alerts
\item Total volume of reported attacks
\item Number of distinct detectors reporting alerts
\item Time since last alert
\item Average and median of intervals between alerts within the history window
\end{itemize}

The first three characteristics can be computed per different time intervals (e.g. the last day and whole history window), each interval resulting in a separate feature.
Another approach is to create time-series of these numbers
(e.g. a number of alerts in each day over the history window)
and use the \ac{EWMA} of the time-series as a feature.
\ac{EWMA}, which is often used as a simple, yet effective, predictor of the next value in a time-series,
can be defined as:

\begin{equation}
\bar{x}_{t} = \alpha x_{t} + (1 - \alpha) \bar{x}_{t-1},
\end{equation}

where $x_t$ is value of the time-series at time interval $t$ and $\alpha \in (0, 1)$ is a smoothing factor, higher values mean more weight is given to recent samples, low values give more weight to older history.

In addition, since we are interested primarily in whether or not there will be an alert in the prediction window, not in the number of alerts or total volume,
it is useful to use \ac{EWMA} of a binary variant of the time-series, which contains $1$ if there are any alerts in a given day, $0$ when there is none.

When the aforementioned features are computed only from alerts reporting the given entity, they only capture temporal characteristics of the entity behavior.
If spatial correlations are expected for the given type of entity, i.e. the behavior of nearby or otherwise similar entities is correlated, an additional set of features can be computed to allow to leverage these correlations.
This set of features is the same as those above, but it counts alerts related to any of the neighboring entities instead of just the evaluated entity itself.
For example, in case of IP addresses, the same set of features can be computed for IP address itself and for the whole /24 prefix it belongs to.

Finally, some of the features reach very high values (e.g. number of alerts or their total volume)
which is not handled well by some machine learning models.
Moreover, it is usually not important whether there was 1000 or 1001 alerts,
but 1 or 2 alerts are a big difference, although the arithmetic difference is the same.
It is therefore recommended to use a log-like nonlinear transformation on most of the features,
which reduces the high values while keeping small differences in small values still significant.
In particular, we recommend to use $log(x + 1)$ for all features meaning a number of something.
For features describing time intervals, $exp(-x)$ should be used instead
to avoid a problem on infinite intervals in case there is no previous alert.
The function maps infinity to 0 and short intervals to values close to 1,
which also makes it consistent with other features that are zero for previously unseen entities and higher for highly active ones.

\subsection{Unbalanced data and recalibration}
\label{sec:subsampling}

Many machine learning models exhibit poor performance when the input data are highly unbalanced,
i.e. when numbers of samples in each class differ significantly.
We expect this will be the case in most applications of our method,
since the entities actually detected as malicious within the prediction window
will be just a small fraction among all evaluated entities.
Therefore, there will be significantly more samples of negative class than those of positive class.

Generally, there are two main approaches to balance a dataset.
A simple and commonly used one is to subsample the majority class.
The other one is to supersample the minority class,
either by duplicating the minority samples, or by creating new artificial samples near the original ones (SMOTE \cite{chawla2002smote}).
Supersampling is more complicated and introduces some drawbacks,
so it is usually chosen only when the dataset is so small that there would be too few samples left for training after subsampling.
This is however not our case. It is usually not an issue to acquire millions of alerts, so we use the subsampling approach.

Subsampling should only be applied on the training dataset,
the testing one should retain the original distribution.
However, this violates the basic machine learning assumption that the training and testing datasets follow the same distribution, resulting in skewed probability estimates (see \cite{dal2015calibrating} for detailed discussion).
Fortunately, as shown in \cite{dal2015calibrating}, this skew can be easily recalibrated by transforming the output of the model learned on subsampled data, $\hat{y}_s$, by the following formula:

\begin{equation}
\hat{y} = \frac{\beta \hat{y}_s}{\beta \hat{y}_s - \hat{y}_s + 1}, \label{eq:recalibration}
\end{equation}

where $\beta = \frac{N^+}{N^-}$ and $N^{+}$, $N^{-}$ represent the number of samples of positive and negative class, respectively, in the original dataset
(assuming negative class is the majority one). 

Alternatively, if an implementation of the selected machine learning model allows weighting of samples,
it is possible to set weight of all negative samples to $\beta$ instead of subsampling. 
However, in the evaluation below, we use the subsampling approach with recalibration
as it means smaller dataset and therefore faster training while results are almost the same.

\section{Evaluation}
\label{sec:evaluation}


In this section we evaluate the general scoring mechanism (introduced in Section~\ref{sec:repscore})
by applying it on real data about network alerts.
In particular, we consider a system as described in Section~\ref{sec:overview}
which receives alerts from various network security tools.
Its NERDS subsystem stores information about reported IPv4 addresses 
and assigns the FMP score to each address based on the estimated probability of receiving another alert related to the same address within the next day.
Our goal is to shed light on how the scoring mechanism is deployed, to confirm and quantify the assumption of repetetive offenders 
and to show prediction results under various settings.
Last but not least, we elaborate a practical application of FMP score -- 
building predictive blacklists for blocking traffic of IP addresses with the highest probability of being malicious.


\subsection{Source of data}

The data used for evaluation come from the Warden system -- an alert sharing tool and community run by CESNET (NREN of the Czech republic).
The alerts are JSON messages reporting on various types of network attacks or other security events.
The attacks and events are detected by various monitoring tools (honeypots, netflow analysis systems, IDS, etc.)
deployed in CESNET and several other networks.
Each alert contains information about at least the time of the event,
its attack category, source IP address(es), identification of the detector
and usually also some measure of attack intensity, like number of connection attempts.
It therefore satisfies all the requirements stated in Section~\ref{sec:formal}.

\subsection{Evaluation setting}

Based on a statistical analysis and our previous experience with the data
(see e.g. a technical report \cite{warden-analysis}),
we decided to set the length of the history window, $w_h$, to 7 days.
The length of the prediction window, $w_p$, is 1 day.
Therefore, we are going to estimate probability of receiving an alert
about a particular IP address within the next 24 hours,
given information about alerts from the previous week.

\subsubsection{Dataset preparation}

For the evaluation, we took three months worth of data from Warden, from September to November 2017.
In total, the dataset consists of $155$ million alerts about $5.3$ million IP addresses, reported by $23$ different sources.
%
The alerts report different types of malicious traffic (attack category) and the vast majority of them are various types of scanning activity, dictionary attacks or exploit attempts.
For the evaluation, we group the alerts into two broad categories
-- port scans (\textit{scan}) and unauthorized access attempts (both dictionary attacks and exploit attempts, \textit{access}).
Therefore for each IP address two FMP scores are computed
-- $\mathit{FMP}_\text{scan}$ predicting alerts of the \textit{scan} category,
and $\mathit{FMP}_\text{access}$ predicting alerts of the \textit{access} category.
Other attack types, such as DDoS attacks,
are reported to Warden only occasionally and are disregarded in this work.

We create the dataset by regurarly sampling the entity database
at 24 different prediction times ($t_0$) within the three months.
At each time $t_0$, we account for only IP addresses that are reported by at least one alert (of given type) within the history window $T_h$ of one week\footnote{Please note that the history windows partially overlap}.
For each such an IP address, a feature vector $\textbf{x}_i$ is computed and a class label $y_i$ is assigned.

We therefore only consider the addresses that has already been reported
and the score thus evaluates the probability they will be reported again.
Theoretically, it is possible to estimate the probability of new occurrence of previously unseen addresses as well,
using information from alerts about other addresses in the same prefix, maliciousness of the ASN and country,
and the supplementary features not based on alerts,
but such scenario is not evaluated in this paper
\footnote{If scoring of previously unseen addresses is needed, we recommend to build a separate model for it. It would be hard to train a single one for both cases due to the extreme imbalance in numbers of addresses observed and not observed in the history window (there are $2^{32}$ IP addresses and just a few millions are known as malicious). Moreover, the model for unobserved addresses can be much simpler, as it can drop some of the input features, which are always zero for such addresses.}.
In total, we got 12.3 million samples related to \textit{scan} alerts (the \textit{scan dataset}),
765,000 samples related to \textit{access} alerts (the \textit{access dataset}).

From each dataset, a random subset of samples is put aside as testing data
(600,000 in \textit{scan} dataset, 100,000 in \textit{access} dataset).
The rest is used for training\footnote{
We keep the training datasets large compared to the test ones since they will be heavily subsampled in the next phase.}.

\subsubsection{Features}

The set of input features computed from the alerts observed in the history window is selected according to
the general guidelines presented in Section \ref{sec:feature-selection}.
For each alert category (\textit{scan} and \textit{access}) the following features are computed for each IP address (taking into account alerts reporting the given IP address):
\begin{enumerate}
\item Number of alerts in the last day
\item Total number of connection attempts (attack volume) in the last day
\item Number of detectors reporting the address in the last day
\item Number of alerts in the last week
\item Total number of connection attempts (attack volume) in the last week
\item Number of detectors reporting the address in the last week
\item \ac{EWMA} of number of alerts per day over the last week
\item \ac{EWMA} of total number of connection attempts per day over the last week
\item \ac{EWMA} of a binary signal expressing presence of an alert ($0$ or $1$) in each day over the last week
\item Time from the last alert (in days)
\item Average interval between alerts within the last week (in days, infinity if less then two alerts were reported)
\item Median of intervals between alerts within the last week (in days, infinity if less then two alerts were reported)
\end{enumerate}

In order to leverage spatial correlations (i.e. the observation that the malicious IP addresses are often close to each other in IP address space), a similar set of features is also computed by taking into account all alerts related to the same /24 prefix as the evaluated IP address.
This \emph{prefix} set contains features 1--9 from the previous list and also two new ones:
\begin{itemize}
\setlength\itemsep{0pt}
\item Number of distinct IP addresses in the prefix reported in the last day
\item Number of distinct IP addresses in the prefix reported in the last week
\end{itemize}


Because there exist significant correlations between scan events and access attempts,
we always use features computed from both \textit{scan} and \textit{access} alert categories (as separate feature sets), regardless of which alert category is to be predicted.

Another two features utilize data about autonomous system numbers (ASN) and geo-location data.
As shown in multiple previous works (see Sec.~\ref{sec:related}), the portion of malicious IP addresses in different countries and ASNs differ significantly.
For each IP address, we determine into which country and ASN it belongs and use the corresponding \emph{maliciousness rates} as input features.
The rate is computed as the number of known malicious IP addresses (i.e. those with at least one alert in the last week) in that country or ASN divided by total number of addresses assigned to that country or ASN.

In total there are 48 alert-based features.

These features are complemented by several features not based on alerts.
All of them are binary, taking a value of $1$ if given condition is met, $0$ otherwise.
First, presence of the IP address on 5 public blacklists\footnote{%
UCEPROTECT, blocklist.de-SSH and Spamhaus PBL, PBL-ISP, XBL-CBL; we also checked several others, but there are almost no overlap in IP addresses with our dataset, which renders them useless for the estimator.},
and a list of dynamic IP ranges\footnote{SORBS DUL} is checked.
Next a hostname associated with the IP address is discovered via a DNS query and several hand-written rules are applied to it.
For example, we search for keywords like “static”, “dynamic”, “dsl”, or look if the IP address is encoded in the hostname.
This results in another 4 features.
All this information is gathered shortly before the prediction time $t_0$.
The reason for estimating whether the IP address is dynamically assigned or not is the expectation that the host behind a dynamic address may change soon after the attack was detected, which intuitively lowers the expectation of repeating attacks from such addresses.

In total, a vector of 58 features is computed for each IP address and prediction time.




\subsubsection{Preprocessing}



The data in both datasets are highly imbalanced. Only 16.5\,\% samples belong to positive class in \textit{scan} data, in \textit{access} data it is only 8.1\,\%.
We therefore apply subsampling of majority class (the \textit{not-detected} one, $y_i = 0$) on the training dataset as described in Section \ref{sec:subsampling}.
This results in 3.88 million training samples in \textit{scan} dataset and 107,000 samples in \textit{access} dataset.

Next, values of most features are non-linearly transformed
as described in section \ref{sec:feature-selection}.
Features expressing number of alerts, connections or detectors are transformed by $log(x+1)$.
Features expressing time intervals are transformed by $exp(-x)$.
Other features are numbers between $0$ and $1$ or binary tags and do not need any transformation.

\subsection{Model fitting}

Subsampled and transformed training data are then passed to a machine learning model to train.
The goal is to minimize the Brier score, i.e. to estimate the probability of the positive class with smallest average error over all samples.
Models are trained separately for \textit{scan} and \textit{access} data.

Finally, the test dataset is passed to each trained model to get the estimated probability of positive class for each sample.
These estimations are transformed by the recalibration formula \ref{eq:recalibration}
and then the results are evaluated.

\subsection{Predictor evaluation}

%
%

First, we show performance of various machine learning models,
then we evaluate impact of various sets of features to see if and how much each of them improves the results.


\subsubsection{Machine learning models}

After an initial study and experimenting with various machine learning methods 
we identified neural networks (NN) and gradient boosted decision trees (GBDT) as the most promising ones.
%
We evaluated many variants of NNs with up to three hidden layers
and several configurations of GBDTs (we used the \textit{xgBoost} implementation \cite{chen2016xgboost}).
Table \ref{tab:scores} shows Brier scores of some of the models for both datasets.

\begin{table}
\centering
\caption{Brier score of different models over testing set of \textit{scan} and \textit{access} datasets}
\label{tab:scores}
\begin{tabular}{|l|c|c|} \hline
                  &   scan & access \\ \hline
NN, 2 layers      & 0.0646 & 0.0549 \\
NN, 3 layers      & 0.0646 & 0.0542 \\
GBDT(100, 3)      & 0.0671 & 0.0529 \\
GBDT(200, 7)      & 0.0628 & 0.0507 \\
\hline
\end{tabular}
\end{table}

The neural networks have 2 or 3 fully connected hidden layers, each with 58 nodes (the number of input features) and rectified linear unit (ReLU) as the activation function.
The output layer is a single node with sigmoid activation.
We also tried different numbers of nodes and activation functions but the results were very similar or worse than the results presented in this Section.

The GBDT models consist of 100 or 200 trees with maximum depth of 3 or 7, respectively.
We tried other combinations as well and the results are not surprising
-- Brier score slowly gets better as model complexity increases but at the same time the training time increases significantly. 

The training time on the \textit{scan} dataset, using 2 CPU cores of an average laptop, takes around 1 hour with \textit{GBDT(200, 7)} model,
while it is below 15 minutes with the simpler GBDT model and the NN based models (training on \textit{access} dataset is finished within a minute for all the models, because the dataset is much smaller).

All Brier scores in Table \ref{tab:scores} are close to zero, meaning a good precision of probability estimation.

\begin{figure}
\centering
\includegraphics[width=8cm]{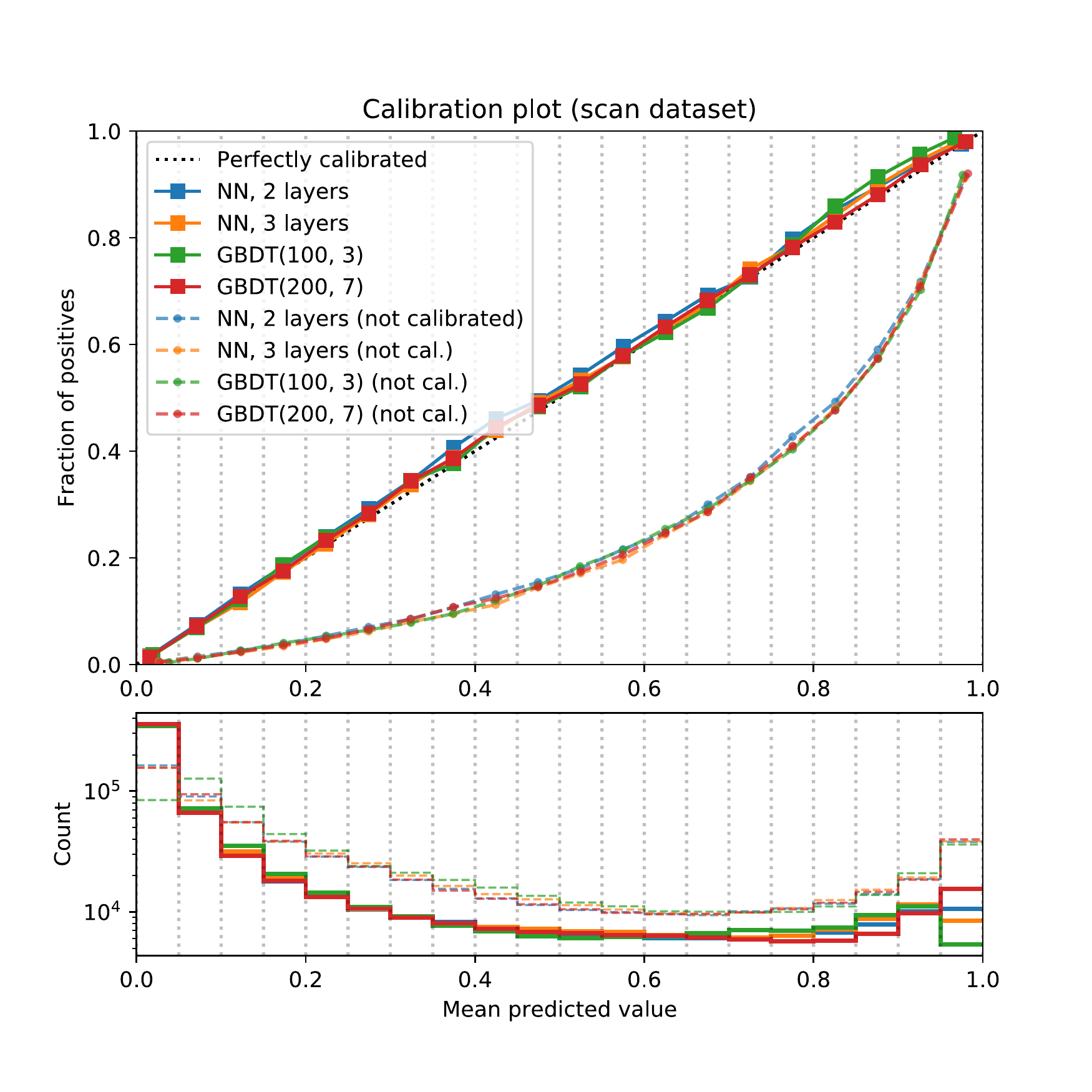}
\caption{Probability calibration curve of 4 different models over the \textit{scan} test dataset.}
\label{fig:scan_cal}
\end{figure}
\begin{figure}
\centering
\includegraphics[width=8cm]{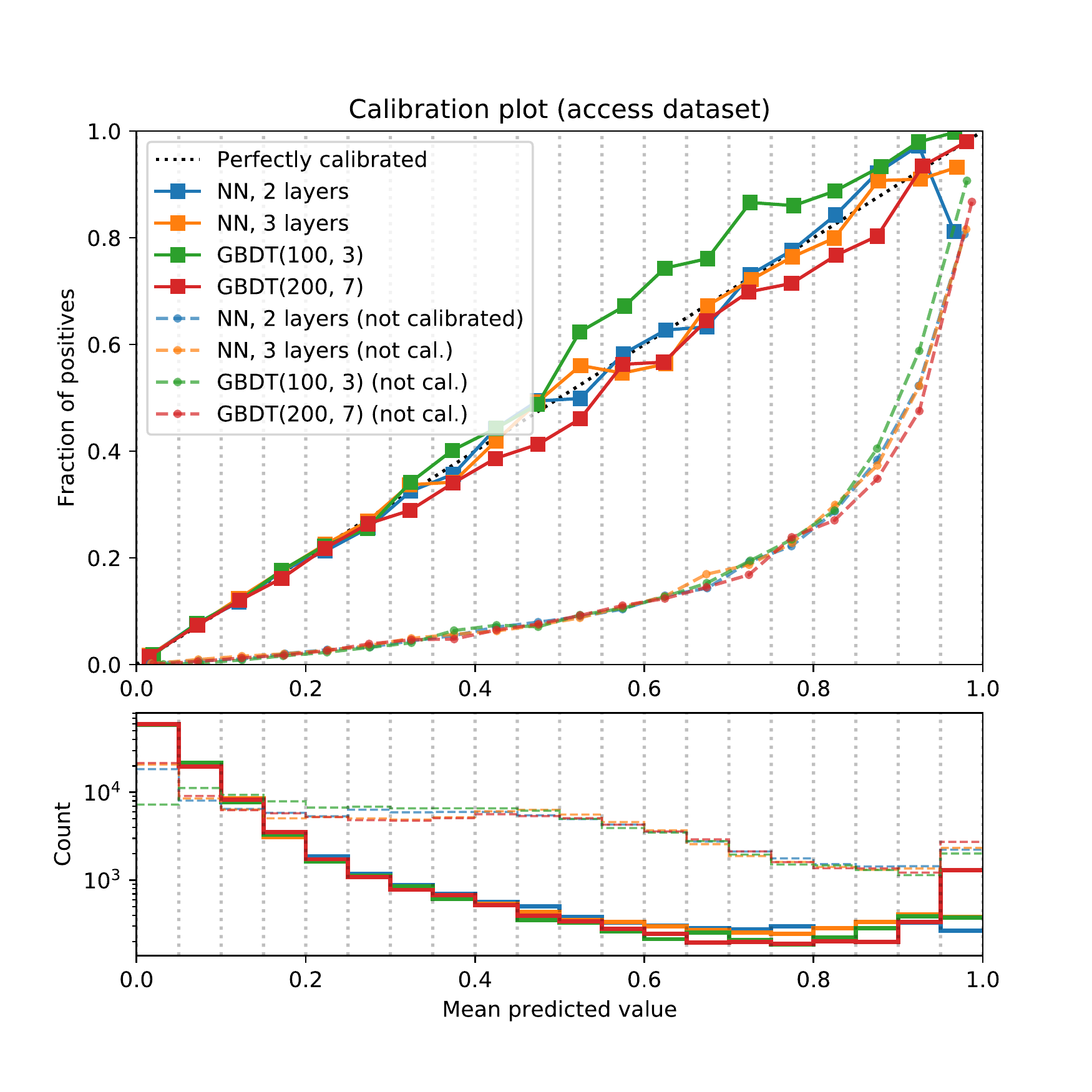}
\caption{Probability calibration curve of 4 different models over the \textit{access} test dataset.}
\label{fig:access_cal}
\end{figure}

A crucial requirement on the FMP score is
that it actually approximates the probability of encountering another alert from the same address.
While this characteristic is already covered by the Brier score,
it is also possible to illustrate is visually.
Figures \ref{fig:scan_cal} and \ref{fig:access_cal} 
show probability calibration curves of all the four models over both datasets.
These curves (sometimes also called reliability curves)
show the distribution of real classes within bins of samples with similar estimated probability of a particular class (the positive one in our case), $\hat{y}_i$.
In other words, samples are binned by their value of $\hat{y}_i$
and for each bin a point is drawn.
Its horizontal position is given by the mean of $\hat{y}_i$ within the bin,
vertical position is equal to the fraction of samples within the bin whose true class is positive ($y_i = 1$).
If the estimator works well, i.e. its output indeed approximates the probability of positive class,
this fraction should be close to the mean of the bin,
and thus the resulting line should be close to the diagonal ($y=x$).

We can see that all models perform very well on the \textit{scan} dataset.
It is slightly worse on \textit{access} dataset, especially at higher values of $\hat{y}$ (approx. between 0.5 and 0.9).
From the histogram below the calibration curves, which shows the number of samples in each bin,
we can see that the number of samples in this range is quite low.
This is mostly because \textit{access} alerts are an order of magnitude less common in our dataset than \textit{scan} alerts.
Nevertheless, the curves are still quite close to the ideal line so we consider estimations of all the models as usable.

To illustrate the importance of recalibration by formula \ref{eq:recalibration},
we also show how the calibration curves look like before the formula is applied
-- showed as dashed lines in Figures \ref{fig:scan_cal} and \ref{fig:access_cal}.
In such case the estimators are highly biased and the estimated probability does not match the real one.
For example, when the uncalibrated estimator outputs 0.6,
the curve shows there is only about 20\,\% chance of seeing an alert within the prediction window, contrary to the expected 60\,\% chance.


In many use cases the FMP score will be used together with 
a threshold (either a fixed value or got by a top-n approach) 
to split IP addresses into “good” and “bad” ones, e.g. to generate a blacklist.
This reduces our binary class probability estimation problem into binary classification. 

It is important to note that the primary goal of our method is not to create a perfect classifier,
as the input data are surely not sufficient for accurate prediction.
This is because behavior of malicious actors is affected by many factors not known to the model,
including such things like random selection of targets in automated scans or attacks.
Therefore, in most cases it is only possible to estimate the probability -- which is our main goal and we evaluated it above.
Nevertheless the metrics used for evaluating binary classification tasks
are generally well understood and can provide further insight into the model performance.

\begin{figure}
\centering
\includegraphics[width=8cm]{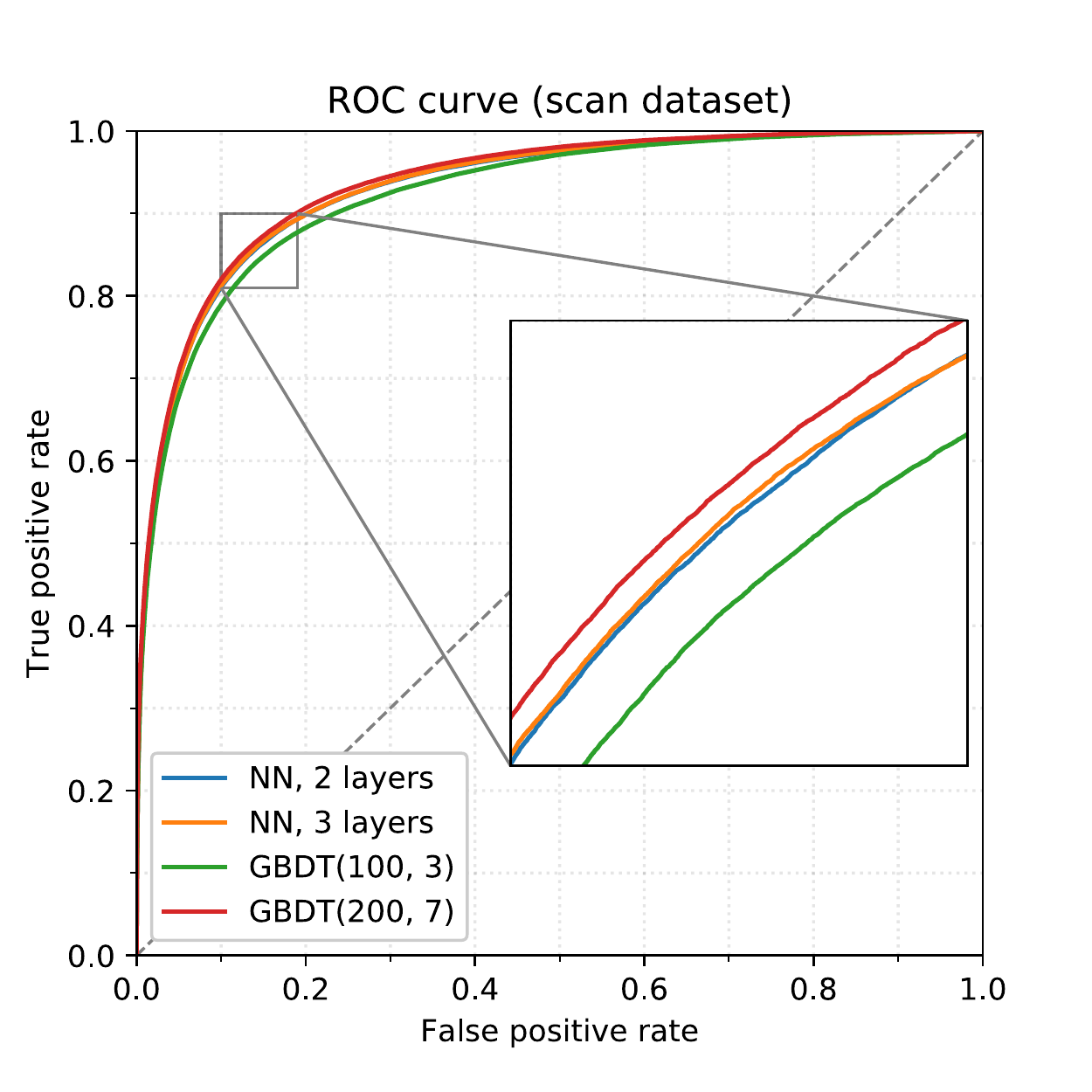}
\caption{ROC curve of 4 different models over the \textit{scan} test dataset after recalibration.}
\label{fig:scan_roc}
\end{figure}

\begin{figure}
\centering
\includegraphics[width=8cm]{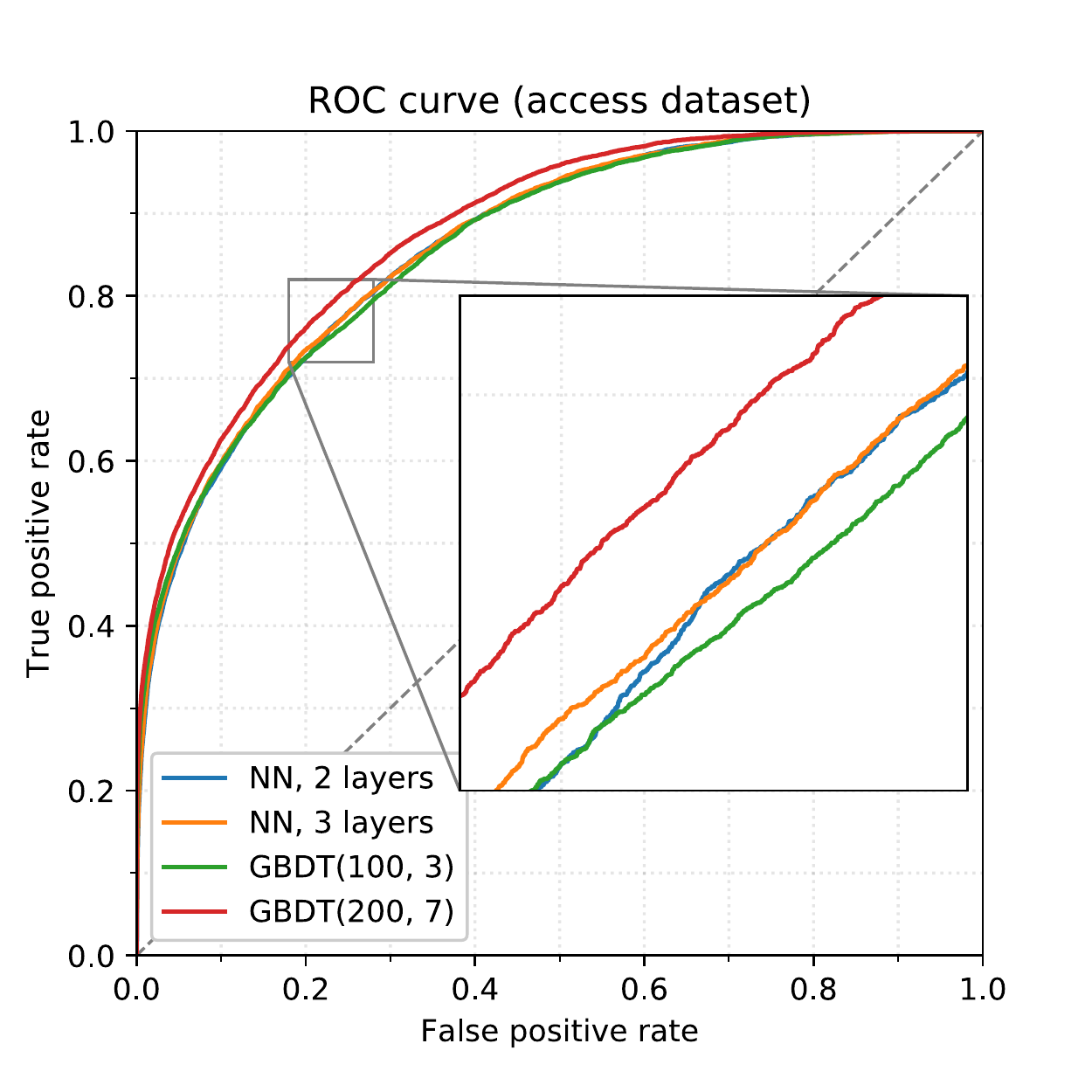}
\caption{ROC curve of 4 different models over the \textit{access} test dataset after recalibration.}
\label{fig:access_roc}
\end{figure}

A common way to visualize results of a binary classification is using Receiver Operating Characteristic (ROC) curves.
These are shown for the evaluated models in Figures \ref{fig:scan_roc} and \ref{fig:access_roc}.
An ROC curve shows the trade-off between false positive and true positive rates
as the value of the threshold changes.
In our case, true positives are the addresses classified as “bad” (i.e. blacklisted) which are then indeed reported as malicious within the prediction window,
false positives are those which are not.
The closer the curve gets to the upper-left corner, the better is the classifier.

All the curves are quite smooth and very similar to each other. The only significant difference is between the datasets,
where scan alerts appear to be more easily predictable than access alerts. 
For example, when the threshold is set to achieve 10\,\% false positive rate,
we can capture over 80\,\% of recurring scanners.
Recall we only included IP addresses that were already reported within the history window.
Also note that false positive here does not necessarily mean blacklisting a legitimate IP address, the address may still be malicious, just not attacking any of the monitored networks within the prediction window. Therefore, it may be just a wasted entry in the blacklist.
This enables us to move the threshold to the area of high false positive rates,
allowing to block almost all recurring attackers
without a significant impact on legitimate traffic.
The only cost is a long blacklist.
More detailed evaluation of the blacklist use case is presented in Section \ref{sec:blacklists}.


Overall, by Brier scores, calibration plots and ROC curves, all evaluated models 
perform very similarly, GBDT being usually slightly better.
Therefore, all further evaluation is performed with only a single model -- \textit{GBDT(200, 7)}.

\subsubsection{Feature sets}

\begin{figure}
\centering
\includegraphics[width=8cm]{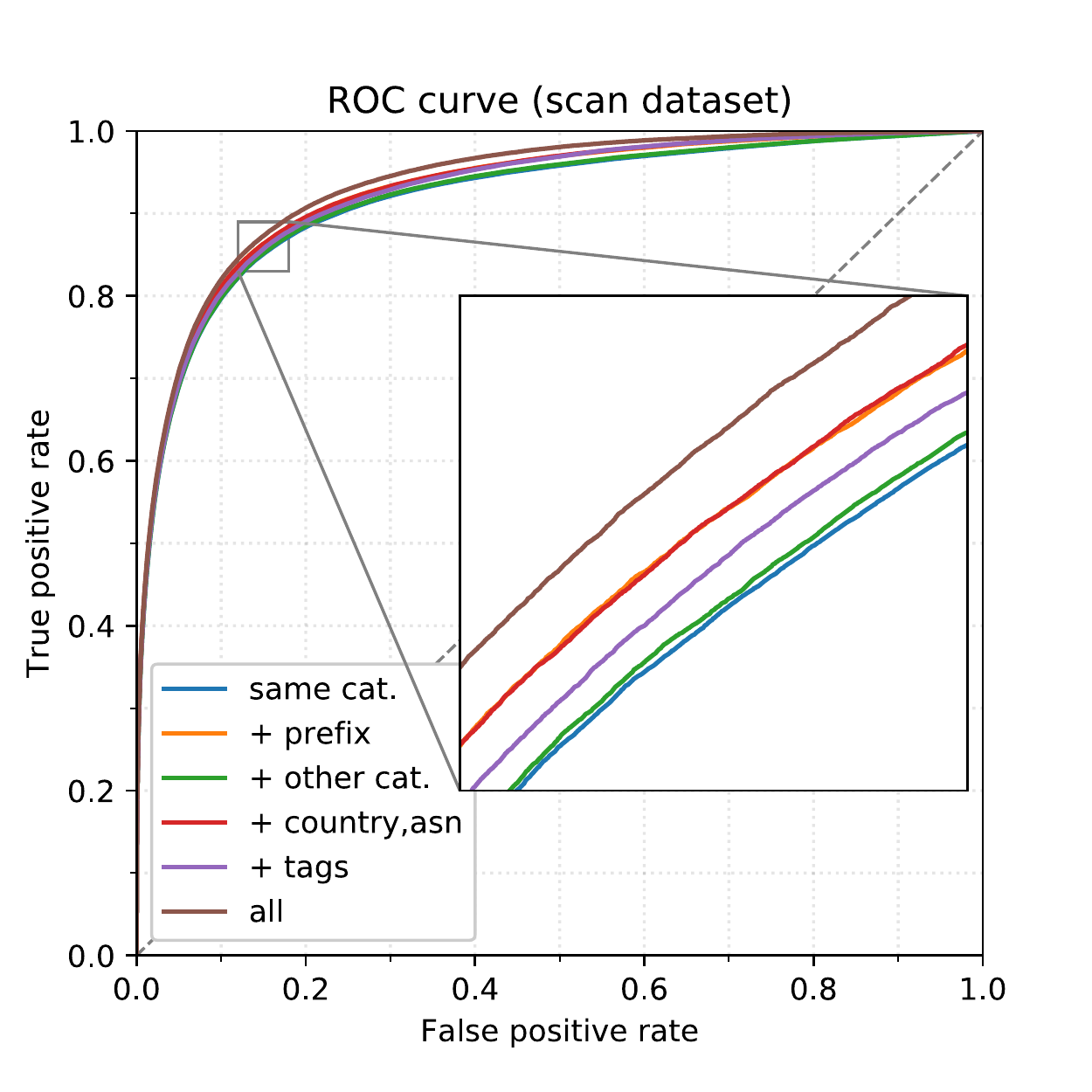}
\caption{ROC curves of different sets of input features (\textit{scan} dataset).}
\label{fig:feature_roc_scan}
\end{figure}

\begin{figure}
\centering
\includegraphics[width=8cm]{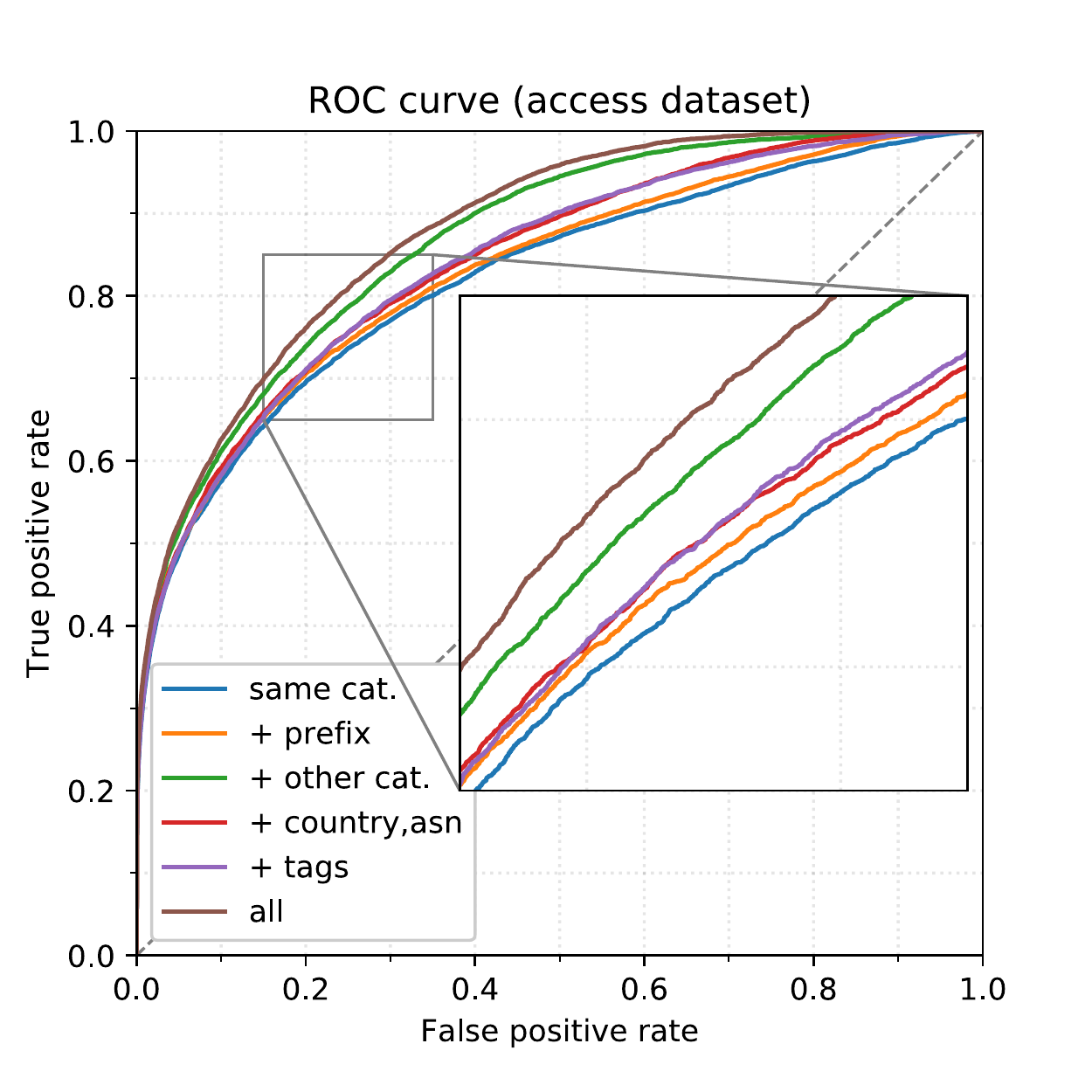}
\caption{ROC curves of different sets of input features (\textit{access} dataset).}
\label{fig:feature_roc_access}
\end{figure}

To see whether all features are indeed useful for prediction,
we evaluate the selected model with different sets of input features.
Every time, the model is trained and tested on the same datasets as in the previous section, just with some of the features removed from feature vectors.
The resulting ROC curves are shown in Figures \ref{fig:feature_roc_scan} and \ref{fig:feature_roc_access}.
The basis is formed by the features computed from alerts of the same category as the one predicted and only about the evaluated address (i.e. not other addresses in the same prefix).
These features, labeled \textit{same cat.}, are always enabled.
We can see that even these basic features provide quite good results on both datasets.
Another four curves show performance of the model when different sets of features
are added to the base one:
features computed from alerts related to IP addresses in the same /24 prefix (labeled as \textit{prefix}),
features computed from alerts of the other category (\textit{other cat.}),
features evaluating rates of malicious IP addresses in the given country and ASN,
and the complementary data not related to alerts (i.e. presence on blacklists and hostname-based tags, labeled here as \textit{tags}).
Finally, the last curve shows the performance when all these feature sets are enabled (i.e. the same as presented in the previous section).

We can observe that all feature sets have measurable effect on the results,
although sometimes very small.
The least useful seems to be the \textit{other category} set of features in \textit{scan} dataset.
That means the alerts of the \textit{access} category are not relevant for prediction of \textit{scan} alerts.
This can be easily explained by the fact that most scanners in our dataset are never reported as performing access attempts,
both because of the character of these attacks (most of \textit{access} attackers also performs scans, but not all scanners try to access the scanned devices)
and the overall disparity in the number of alerts of those categories in our dataset.
Indeed, when we look at ROC curves of the \textit{access} dataset,
the alerts of the \textit{other category} (i.e. \textit{scan}) improves the results very significantly.

A similar but reverse effect can be observed with the \textit{prefix} features.
Addition of features computed from alerts of the whole /24 prefix improves the results significantly in the \textit{scan} dataset,
but there is only minor improvement in the \textit{access} dataset.
We explain this by the lower number of \text{access} attack sources in our dataset, which means there is a lower chance of observing many such addresses in the same prefix, so the predictor can rarely use this type of correlation.

Another set of features utilizing spatial correlations, the ones based on geo-location and ASN data,
show almost the same effect as the \textit{prefix} features on the scan dataset.
On the access dataset, they provide slightly better results, probably because the grouping of addresses based on country and ASN provides much larger groups than /24 prefixes, so there is higher chance there are multiple attacking addresses in the same group.

The last set of features are the \textit{tags} obtained from supplementary, non-alert data.
We can see that in both cases presence of these features improves the results significantly.
Unsurprisingly, combination of all the features provides better results than any of the feature sets alone.
Overall, we can conclude that all of the feature sets prove to be useful.

\subsection{Using FMP score to create predictive blacklists}
\label{sec:blacklists}



%
%
%

In this section we evaluate one of the possible use cases of the scoring method
-- generating blacklists of a user defined size.
In this use case a list of IP addresses with the highest FMP score (a blacklist) is created at the end of each day
and used to block traffic\footnote{%
Or apply rate limiting or any other restrictive measures, depending on user's needs.}
from these addresses during the next day.

The size and restrictiveness of the blacklist can be controlled by the user
-- either by taking a fixed number of the worst IP addresses,
or taking all IP addresses with FMP score greater than a fixed threshold.
Assuming the probability estimation is accurate,
it is guaranteed that such blacklist has the highest hit count possible with the given length of the list.
Following \cite{zhang2008hpb, soldo2011blacklisting}, we define \textit{hit count} as the number of IP addresses on the blacklist that are correctly predicted, i.e. the IP is indeed detected and reported by an alert within the prediction window.
We also define \textit{hit rate}, which is hit count divided by the size of the list. 

In this section we evaluate the effectiveness of blacklists generated using FMP scores.

\begin{figure}
\centering
\subfloat{\includegraphics[width=6cm]{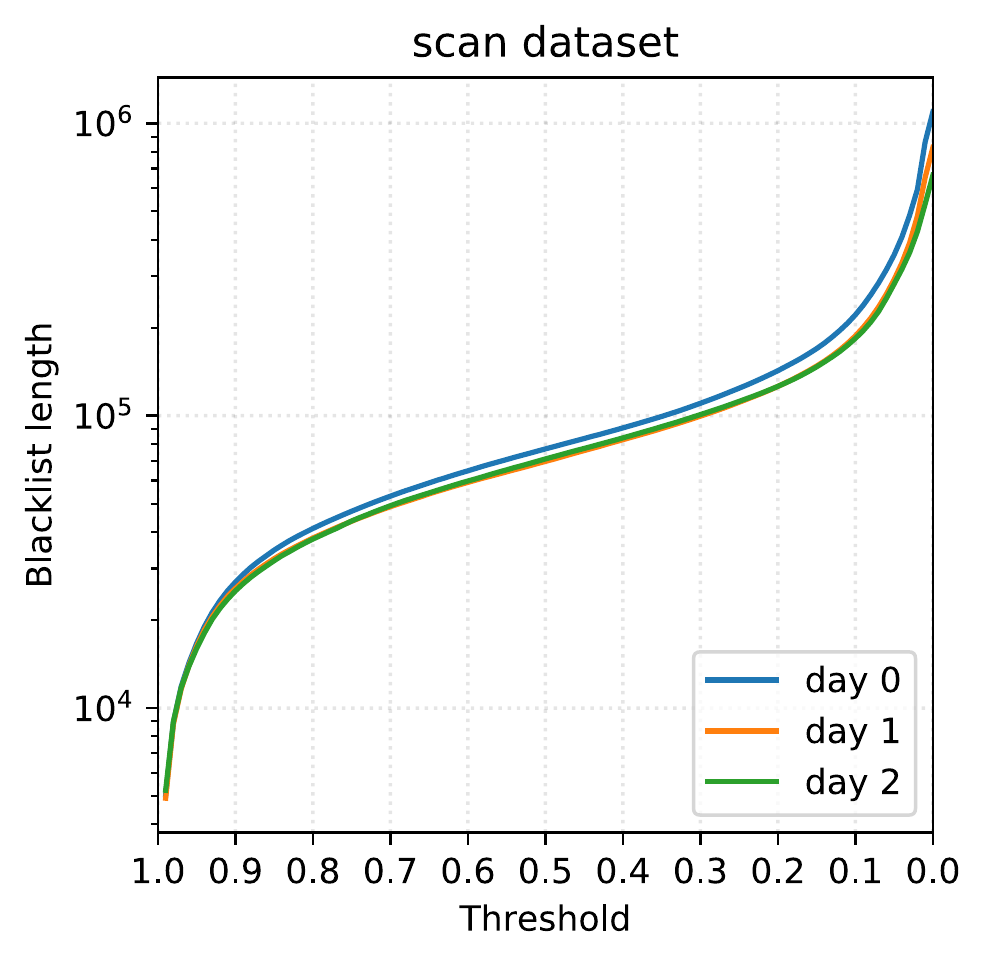}}
\subfloat{\includegraphics[width=6cm]{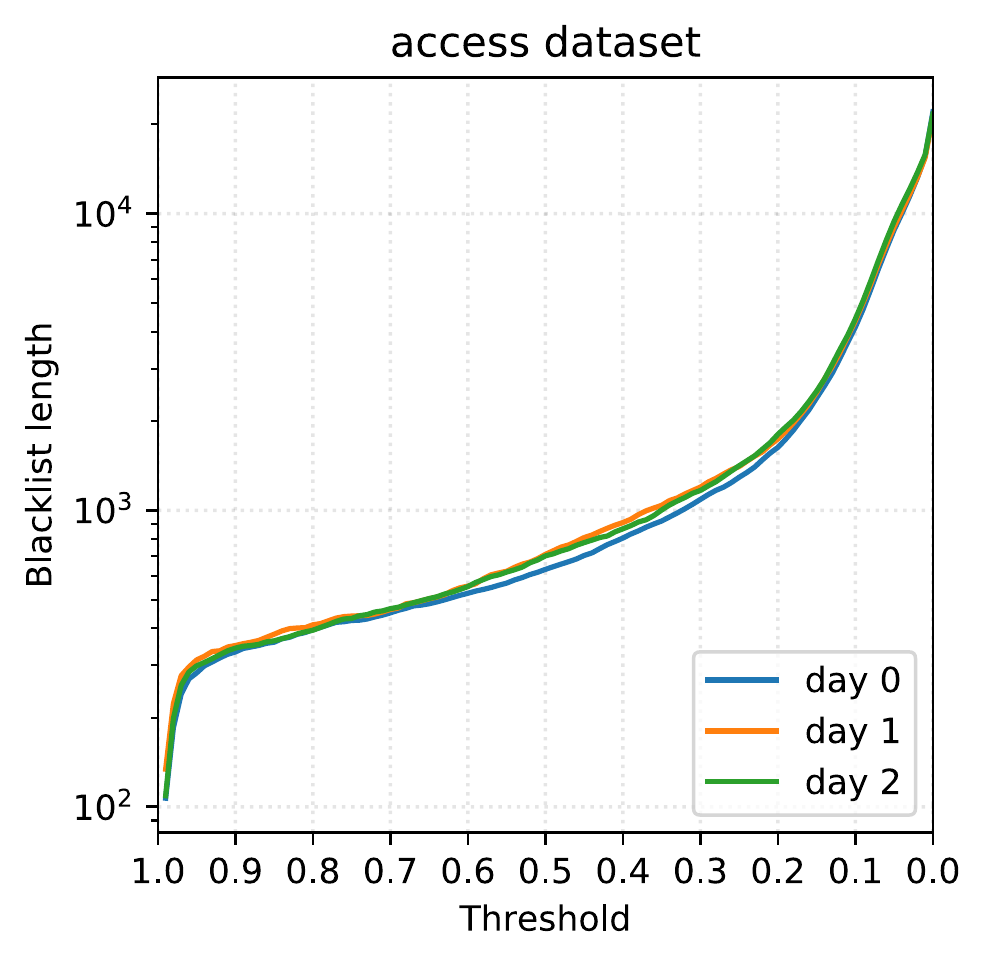}}
\caption{Blacklist size as a function of threshold value applied on FMP score. Each line corresponds to the blacklist generated for one of the three days.}
\label{fig:blacklist-size}
\end{figure}

We took data from three days in the first half of December 2017, i.e. shortly after the data used for training.
For each day, we computed feature vectors of all addresses reported within the previous week
and assigned them FMP score using the estimator trained in the previous section (\textit{GBDT(200,7)} with all features).
Then, we generated a list of IP addresses for each day, sorted by FMP score in decreasing order.
Blacklists are generated from these lists by taking the first $N$ entries or all entries with FMP score greater than or equal to a fixed threshold. 
Figure \ref{fig:blacklist-size} shows the relationship between the threshold and the size of the list
for each of the days and datasets.

Further, we only use the \textit{access} dataset,
since unauthorized access attempts are more severe events than ordinary scans and it makes more sense to block them
or apply some strict rules on the related traffic.

We evaluate hit count of FMP-based blacklists of different sizes $N$
and compare them to blacklists created in a more traditional way.

As a baseline we created blacklists from the same data (i.e. alerts from Warden) but using a basic method
-- listing the most active attackers reported by all the detectors contributing to the alert sharing system within a history window
(called GWOL, \textit{global worst offender list}, in \cite{zhang2008hpb, soldo2011blacklisting}).
We generate these lists using two different lengths of the history window, one day ($\mathit{GWOL}_1$) and 7 days ($\mathit{GWOL}_7$).
Similarly to our approach, the GWOL lists can be generated with any number of entries,
so we always compare an FMP-based list and GWOL of the same length.

We also compare these lists with three real third-party blacklists, namely
UCEPROTECT, blocklist.de--SSH (\textit{bl.de-ssh}) and BruteForceBlocker\footnote{\url{http://danger.rulez.sk/index.php/bruteforceblocker/}} (\textit{bfb}).
These lists have fixed size and are based on different input data.

\begin{table}
\centering
\caption{Hit count, hit rate and fraction of attackers blocked by blacklists of different sizes}
\label{tab:blacklists}
\begin{tabular}{|r|c|r|r|r|r|} \hline
\multicolumn{1}{|c|}{$N$} & blacklist & \multicolumn{1}{|c|}{$T$} &  hit count  &  hit rate  & \% of attackers \\ \hhline{|-|-|-|-|-|-|}
        & FMP        &  0.99 &      100    &     100\,\%  &   2.3\,\%  \\
100     & GWOL$_1$   &    -- &       83    &      83\,\%  &   1.9\,\%  \\ 
        & GWOL$_7$   &    -- &       71    &      71\,\%  &   1.6\,\%  \\ \hline 
        & FMP        &  0.68 &      443    &      89\,\%  &  10.1\,\%  \\  
500     & GWOL$_1$   &    -- &      236    &      47\,\%  &   5.4\,\%  \\ 
        & GWOL$_7$   &    -- &      233    &      47\,\%  &   5.3\,\%  \\ \hline 
        & FMP        &  0.18 &      862    &      43\,\%  &  19.7\,\%  \\ 
2000    & GWOL$_1$   &    -- &      650    &      33\,\%  &  14.9\,\%  \\ 
        & GWOL$_7$   &    -- &      579    &      29\,\%  &  13.2\,\%  \\ \hhline{|-|-|-|-|-|-|}
388444  & uceprotect &    -- &      463    &    0.12\,\%  &  10.6\,\%  \\
8063    & bl.de-ssh  &    -- &      336    &     4.2\,\%  &   7.2\,\%  \\
1503    & bfb        &    -- &       70    &     4.7\,\%  &   1.6\,\%  \\ \hline
\end{tabular}
\end{table}

Table \ref{tab:blacklists} shows performance of all the blacklists.
The FMP-based and GWOL blacklists are generated by taking top $N$ IP addresses, for $N$ being 100, 500 and 2000.
The column labeled as $T$ shows which FMP threshold corresponds to given size of the blacklist.
In other words, the list contains all IP addresses with $\mathit{FMP}_\text{access} \geq T$.
The hit count column shows the number of addresses attacking in a given day that would be blocked by the blacklist (number of hits).
Hit rate is simply hit count divided by $N$. It shows proportion of the blacklist that was indeed used to block some attacks.
All numbers in the table are averages over the three days.

Generally, smaller blacklists have higher hit rate, which is expected since they contain IP addresses with the highest probability of future alerts (or most active ones in the past in case of GWOL).
The FMP-based blacklist with 100 entries is especially efficient as all of the listed addresses indeed attacked in two of the days, in the third it was 99.
In all cases, the FMP-based blacklists are significantly more efficient than any of the GWOL ones by means of hit count and hit rate.

On average, there were 4,376 distinct attacking IP addresses in each day.
The last column shows how many of these attackers were blocked by each blacklist.
There it is important to note that around 60\,\% of attackers in each day are “new”, i.e. they has never been detected in the previous week,
so their attacks are almost impossible to predict.
Achievable maximum of the fraction of predicted attackers is therefore 40\,\%.
None of the blacklists get close to this maximum, but still, the FMP blacklists are significantly better than the others.

The third-party blacklists prove to be very inefficient by means of hit rate,
as only a small portion of listed addresses are observed by detectors in Warden.
This is given by different sources of data used to build these blacklists,
so they also list many attackers that do not target any of the networks contributing to Warden.
Nevertheless, if large size of a blacklist is not an issue, these lists can be used to complement the FMP-based one.
Indeed, a combined list of FMP-based list thresholded at 0.5 (689 entries) and the three third-party blacklists (397250 entries in total) can block 24.2\,\% of attacks.
However, the hit rate is only 0.26\,\%, meaning that vast majority of entries are unused.
Also, too large blacklists increase the chance of blocking a legitimate traffic,
so the smaller, more efficient FMP-based blacklists may be preferred in many cases.

\section{Conclusion}\label{sec:conclusion}
In this article, we introduce the \acf{NERDS}, a system that is intended for being part of \acp{CIDS} and collaborative defenses to assist with the prediction of future attacks and with prioritizing the alert data. 
We defined the \acf{FMP}, a score that evaluates network entities by predicting their future behavior, and proposed a method to create the predictor by utilizing the machine learning techniques.

Evaluation of the predictor on a real dataset, containing two types of alerts reporting malicious IP addresses, demonstrates that our proposed method is effective. Additionally, the \ac{FMP} score can be used both for ranking IP addresses (enables alert prioritization) as well as for predicting a set of addresses, that will most probably attack on the next day.
Furthermore, we demonstrate that the \ac{FMP} score can be used to generate predictive blacklists. Their efficiency is measured as the number of listed attackers relative to blacklist size. Our results show that \ac{FMP}-assisted blacklists clearly outperform the traditional ones.

Another possible use case, evaluated in a separate work~\cite{jansky2018augmented}, is using FMP score of suspicious IP addresses as one of the criteria for separating malicious and legitimate traffic in a DDoS mitigation algorithm. In this method, traffic from IP addresses with high FMP score have higher chance to get blocked during a DDoS attack.

With regard to future work, we are currently introducing \ac{NERDS} and the \ac{FMP} score into \textit{PROTECTIVE}, a system for cyber threat intelligence sharing and analysis being developed by a consortium of $10$ academic and commercial partners from Europe.
We are also exploring a possibility of combining the FMP score, as an indicator of malicious activities, with information about normal traffic in a network (from NetFlow data) to improve the precision of the blacklists and lower the chance of blocking a legitimate traffic.
Lastly, we plan to try to use deep learning methods to further improve the prediction.
Some of these methods could allow to predict not only probability of future alerts, but also some of their parameters, like type of attack, expected intensity, or the target.

\section*{Acknowledgments}
This work has received funding from the European Union’s Horizon 2020 Research and Innovation Program, PROTECTIVE, under Grant Agreement No 700071.

\section*{References}

\bibliography{bib/Cybersecurity,bib/manual}

\end{document}